# Effects of high pressure on the physical properties of MgB$_2$.

**T. Prikhna**[1], W. Gawalek[2], Ya. Savchuk[1], A. Soldatov[3], V. Sokolovsky[4], M. Eisterer[5], ,
H.W. Weber[5], J. Noudem[6], M. Serga[1], V. Turkevich[1], M. Tompsic[7], V. Tkach[1], N. Danilenko[8],
W. Goldacker[9], F. Karau[10], I. Fesenko[1], M. Rindfleisch[7], J. Dellith[2], M. Wendt[2], S. You[3],
V. Meerovich[4], S. Dub[1], V. Moshchil[1], N. Sergienko[1], A. Kozyrev[1], T. Habisreuther[2],
Ch. Schmidt[2], D. Litzkendorf[2], P. Nagorny[1], V. Sverdun[1].

[1] *Institute for Superhard Materials of the National Academy of Sciences of Ukraine, Kiev 04074, Ukraine, prikhna@mail.ru, prikhna@iptelecom.net.ua*

[2] *Institut für Photonische Technologien, Jena, D-07745, Germany*

[3] *Luleå University of Technology, Department of Applied Physics & Mechanical Engineering, SE-971 87 Luleå, Sweden*

[4] *Ben-Gurion University of the Negev, P.O.B. 653, Beer-Sheva 8410,5 Israel*

[5] *Vienna University of Technology, Atominstitut, 1020 Vienna, Austria*

[6] *CNRS/CRISMAT, 6, Bd du Maréchal Juin, CNRS UMR 6508, 14050, Caen, France*

[7] *Hyper Tech Research, Inc. 1275 Kinnear Road Columbus, OH 43212, USA*

[8] *Institute for Problems in Material Science of the National Academy of Sciences of Ukraine, 3 Krzhizhanovsky Street, Kiev, 03680, Ukraine*

[9] *Forschungszentrum Karlsruhe, Institut für Technische Physik, 3640, D-76021, Karlsruhe, Germany*

[10] *H.C. Starck GmbH, Goslar 38642, Germany*

E-mail: prikhna@iptelecom.net.ua, prikhna@mail.ru

**Abstract**
The synthesis of MgB$_2$-based materials under high pressure gave the possibility to suppress the evaporation of magnesium and to obtain near theoretically dense nanograined structures with high superconducting, thermal conducting, and mechanical characteristics: critical current densities of 1.8–1.0·10$^6$ A/cm$^2$ in the self field and 10$^3$ A/cm$^2$ in a magnetic field of 8 T at 20 K, 5-3·10$^5$ A/cm$^2$ in self field at 30 K, the corresponding critical fields being $H_{C2}$ = 15 T at 22 K and irreversible fields $H_{irr}$ =13 T at 20 K, and $H_{irr}$ =3.5 T at 30 K, thermal conduction of 53±2 W/(m·K), the Vickers hardness H$_v$=10.12±0.2 GPa under a load of 148.8 N and the fracture toughness K$_{1C}$ = 7.6± 2.0 MPa m$^{0.5}$ under the same load, the Young modulus E=213 GPa. Estimation of quenching current and AC losses allowed the conclusion that high-pressure-prepared materials are promising for application in transformer-type fault current limiters working at 20-30 K.

1. INTRODUCTION

Interest in the MgB$_2$-based materials for practical application despite the comparatively low superconducting (SC) transition temperature, can be explained on the one hand, by the transparency of clean grain boundaries to current [1, 2], and the possibility to achieve high critical magnetic fields and critical current densities in polycrystalline material, which in turn leads to the simpler and cheaper preparation technique (than in the case of high-temperature superconductors), by aptitude of the materials for both the large scale applications and the electronic devices. On the other hand, this interest can be explained by the intensive development of technologies that use liquid hydrogen as an alternative fuel for motors, water and aviation transports as well as for cooling long power transmission lines, because the boiling temperature of liquid hydrogen (20 K) can be the working temperature for MgB$_2$-based superconductive materials. For the efficient application of MgB$_2$-based superconductors a further improvement of the in-field $J$c is of great importance.

There are two classes of properties that can be responsible for the $J_c$ and might be termed 'intrinsic' and 'extrinsic' [3]. According to classification given in [3], intrinsic (i.e. intragranular) properties of polycrystalline MgB$_2$ are the upper critical magnetic field, $H_{c2}$ and flux pinning (or, alternatively, intragrain $J_c$). An enhancement of the extrinsic properties (connectivity and porosity) can be attained by substantial across-the-board increases in $J_c$ and can accompany an increase in the effective superconducting cross-sectional area for the conduction of transport current. Any kind of disorder potentially changes the MgB$_2$ superconducting properties. Disorder can be introduced in a controlled way by doping or irradiation, but often arises from the preparation conditions and generally decreases the transition temperature. The connections between the grains remain weak, since dirty grain boundaries potentially reduce the critical currents. It was concluded that grain boundaries are the dominant pinning centres in thin films, where high critical current densities close to 15% of the depairing current density were reached [2]. According to this, the expected maximum for loss-free supercurrents is achieved in the case of the optimized pinning when nanosized $MgB_2$ grains provide enough grain boundaries to reach the theoretical limit. The observations show that grain boundary pinning may be dominant in consolidated $MgB_2$ [4]. The bad connectivity between grains reduces the critical currents by a factor of about five in today's best 'high field superconductors'. The anisotropy further suppresses the critical currents at high magnetic fields and is successfully described by a percolation model. The introduction of disorder increases the upper critical field and reduces its anisotropy, leading to higher currents at high magnetic fields [2].

The introduction of dopants can have one or more effects, which may result in an increase in the bulk pinning strength [3]: (1) by increasing crystal's $H_{c2}$ and $H_{irr}$, (2) by forming a wide distribution of point pinning centers and (3) by producing localized lattice strains, which also contribute to flux pinning.

Pressure is of great importance in the manufacture of superconducting $MgB_2$-based bulk materials. Pressure allows suppressing a volatility of Mg, impeding its oxidation, and promoting the formation of a mechanically stable denser structure. A denser material usually exhibits higher superconducting properties; it is more stable against degradation during exploitation, less reacts with a moisture, etc. Even in the case of pressureless synthesis (in-situ) or sintering (ex-situ) a preliminarily densification (compaction), i.e. pressure treatment, plays a great role in attaining high superconducting and mechanical characteristics of the produced materials. Recently the efficiency of high-pressure densification has been shown for wire manufacturing process, in particular [5].

Despite the comparatively simple lattice structure of $MgB_2$, to find correlations between the $MgB_2$-based material structural features and its superconducting properties is a very complicated task. It can be explained by a considerable difficulty that arises from the electronic structure, by the necessity to detect the amount and distribution of light element boron, to analyze the boron-containing compounds [6] in nanostructural materials, which are often porous and in addition, can easily react with oxygen and hydrogen.

In this paper we present the results of studying the structure, superconducting, thermophysical, and mechanical properties of the $MgB_2$-based materials prepared by us *in-situ* from Mg and B or *ex-situ* from $MgB_2$ by different pressure-using techniques: high pressures (2 GPa), hot pressing (30 MPa), spark plasma sintering (50 MPa), and isostatic pressing (0.1 GPa) at 600–1200 $^o$C. In the describing of correlations between structure and critical current density the emphasis is given to high-pressure synthesized materials (without additions and with addition of Ti, Ta, Zr, SiC) having higher superconducting and mechanical properties and high density. The influence of deviation from the stoichiometry of $MgB_2$ up to $MgB_{20}$ in the initial mixture of Mg and B on superconducting properties of high-pressure-synthesized materials is considered. The quenching current and AC losses measured by inductive methods using rings, cut from materials synthesized at high-pressure and hot pressing, are presented and analyzed from the point of view of the materials applicability in transformer-type fault current limiters.

## 2. EXPERIMENTAL

Samples were prepared using (1) high pressures, HP, in recessed-anvil high-pressure apparatuses, , (under 2 GPa) [7], (2) hot pressing, Hot-P, ( under 30 MPa), (3) spark plasma sintering, SPS, (under 50 MPa), (4) hot isostatic pressing, HIP, (under 0.1 GPa gas pressure) and (5) pressureless synthesis, PL, (under 0.1 MPa of Ar). In the case of HP, Hot-P, SPS the cylindrical or rectangular blocks were produced from a mixture of Mg and B or $MgB_2$ with and without additions. In the case of HIP and PL the ring-shaped items were manufactured from a mixture of Mg and B. The formation of rings were performed using broaching: the initial mixture of powders was compacted between coaxial steel tubes by pushing the broach tool inside the hole of the inner steel tube while around the outer steel tube the compressive pressure was created. During the synthesis or sintering the samples were in contact with hexagonal BN (HP, Hot-P), steel (HIP, PL) or graphite (SPS). The synthesis and sintering processes were studied in the 600–1200 °C temperature range. As the initial materials we used powders of $MgB_2$ (I) with an average grain size of 10 μm and 0.8 % of O (H.C. Starck), $MgB_2$ (II) of 98 % purity (Alfa Aesar), and $MgB_2$ (III) with an average grain size of 4.2 μm, 1.9% of O (H.C. Starck), and several types of amorphous boron: B(I) with grains <5 μm, 0.66 % O (H.C. Starck), B(II) (HyperTech, USA), B(III) with grains of 4 μm, 1.5 % O (H.C. Starck), B(IV) of 95-97% purity, grains of 0.8 μm, 1.7 % of O (MaTecK), and B(V) 1.4 μm, 1.9% O (H.C. Starck); metal magnesium chips (I) (Technical Specifications of Ukraine 48-10-93-88), powder of Mg (II) produced by HyperTech (USA), Ti (of size 1-3 μm, MaTecK, 99% purity), Ta (technical specifications 95-318-75, 1-3 μm), Zr (of 2 - 5 μm, MaTecK, 94-98% purity) or SiC (200–800 nm, H.C. Starck). To produce $MgB_2$-based materials, metal magnesium turnings and amorphous boron were taken in the stoichiometric ratio of $MgB_2$. To study the influence of Ta, Zr, Ti or SiC, the powders were added to the stoichiometric $MgB_2$ mixture in the amount of 10 wt%. The components were mixed and milled in a high-speed activator with steel balls for 1-3 min and then tabletized. To study the processes of the higher borides formation, Mg and B (III) were taken in the $MgB_4$-$MgB_{20}$ stoichiometry and heated up to 800–1200 °C at 2 and 5.5 GPa for 1h.

The structure of the materials was analyzed using TEM, SEM, and X-ray diffraction. A ZEISS EVO 50XVP scanning electron microscope (resolution of 2 nm at 30 kV), equipped with: (1) an INCA 450 energy-dispersion analyzer of X-ray spectrums (OXFORD, England), which allows one to perform the quantitative analysis from boron to uranium with a sensitivity of 0.1 wt %; microprobe is 2 nm in diameter; (2) a HKL Canell 5 detector of backscattering electrons (OXFORD, England), which can give (using the Kikuchi method) the diffraction reflections of electrons from 10-1000 nm areas and layers were employed. For the SEM study a JXA 88002 microscope was used as well. The microstructure analysis on the nanometer scale was carried out using JEM-2100F TEM equipped with an Oxford INCA energy detector. The quantitative TEM-EDX analysis of boron was performed using the Oxford INCA energy program, microprobe is 0.7 nm diameter.

The values of $j_c$ were estimated by an Oxford Instruments 3001 vibrating sample magnetometer (VSM) using Bean's model. Magnetization measurements were also carried out on Quantum Design PPMS equipped with vibrating sample magnetometer attachment. The thermal conductivity at room temperature was measured using an IT3-MXTI special device for determination of a thermal conductivity coefficient by a nonstationary method. For Raman studies, we used a WiTec CRM-200 confocal imaging system with the HeNe laser excitation (photon energy of 1.96 eV). The laser beam was focused on the sample using 20x or 100x objectives. The spectra were collected in back scattering geometry with a resolution of 2 cm$^{-1}$. Incident laser power was measured directly on the sample stage and did not exceed 2 mW in order to avoid sample heating.

Hardness was measured using a Mod. MXT-70 Matsuzawa microhardness tester, $H_V$ (using a Vickers indenter) and a Nano-Indenter II, $H_B$ (using a Berkovich indenter).

The transport critical current and AC losses were measured by the inductive methods (see [8, 9] using two MgB$_2$ rings cut out from bulk cylinders using a comparatively simple cutting method developed by us. The method is nondestructive for material superconducting properties and allows easily and precisely cutting brittle and reactive with water $MgB_2$-based materials (even highly dense HP-synthesized ones). The rings were tested as elements of inductive fault current limiter models to investigate the influence of the connection between grains on the quenching current as well as to determine the possibility of $MgB_2$-based material application for current limitation.

3. RESULTS AND DISCUSSIONS

3.1 Superconducting characteristics of the materials produced under different pressures.

Figure 1 shows the dependence of the critical current density, $j_c$, on the magnetic field at 20 and 30 K measured by the magnetization method for the materials fabricated by different pressure-using methods. The material's densities, $\gamma$, which were attained, are presented in Table 1. The highest $\gamma$ and $j_c$ are demonstrated by materials prepared at a pressure of 2 GPa. The currently available high-pressure apparatuses (HPA) allow producing cylindrical blocks up to 62 mm in diameter and 20 mm in height under 2 GPa pressure and HPA for manufacturing blocks up to 150 mm in diameter is being constructed.

Figures 2a, b demonstrate the critical current density, $j_c$, vs. magnetic field at 20 K of the high pressure-synthesized materials with and without additions produced from two different types of boron at the optimal temperatures from the point of view of highest $j_c$ in low and high magnetic fields. At 20 K the high-pressure-synthesized (*in situ*) $MgB_2$-based materials 1HP (with 10 wt.% of SiC additions) and 2HP (without additions) prepared from boron (I) and magnesium (I) demonstrated the highest $j_c$ in  magnetic fields up to  5.5 and 4.5 T, respectively, while the $j_c$ of 4HP material synthesized without additions from boron (II) and fine magnesium (II) was the highest in the fields above 5.5 T, Fig. 1a. The experiments show that for 4HP material (without any additions) at 18.5 K the irreversible field was $H_{irr}$=15 T (Fig. 3a) and at 22 K the upper critical field was $H_{c2}$=15 T (Fig.3b) which are the highest values ever mentioned in the literature. For comparison the $H_{c2}$ = 14 T exhibited by the carbon-doped magnesium diboride $MgB_{2-x}C_x$ (x=0.1) at 18.5 K [10], while for 4HP such $H_{c2}$ was observed at 23 K. Since our measurements can be carried out only up to 15 T, it was not possible to experimentally find $H_{c2}$ at a temperature lower than 22 K.

The highest values of $j_c$ at 30 K in the fields below 1 T were obtained for the samples high-pressure synthesized in-situ without additions (Fig. 1 b) that can be attributable to their higher transition temperature. The transition temperature, $T_c$, of the samples with SiC additions prepared from the same initial boron under the same pressure-temperature-time conditions was somewhat lower as compared with that of the material without additions and this was due to an increase of the amount of nonsuperconducting phase in the superconducting matrix. Unexpectedly high (ever observed) upper critical and irreversible fields as well as critical current density in magnetic fields higher than 5.5 T (Figs. 1 and 3, curves 5) are demonstrated by rather porous HP4 material (Fig. 4) high-pressure–synthesized from fine magnesium (II) and boron (II) at a very low temperature (600 °C). But this material has a low transition temperature of 34.5–34 K (Fig. 5, curve 8 ) and low values of critical current density in low magnetic fields (Figs. 1a, b). Attempts to increase the $j_c$ value of the material by increasing the synthesis temperature (under the same pressure-time conditions) failed: higher synthesis temperature resulted in a decrease of the material's $j_c$ in the 10–35 K temperature and 0–10 T magnetic field ranges.

As it possible to see from the results shown in Fig. 2 such additions as SiC and Ti may increase $j_c$ of high-pressure-synthesized $MgB_2$-based materials at least in some magnetic fields, but the absolute values of $j_c$ depends on the type of initial boron and synthesis temperatures $T_s$. Usually the materials synthesized at high pressure and temperature (1050 $^o$C) exhibit high $j_c$ in zero, low, and medium magnetic fields (up to 4.5–7 T at 20 K), while the materials synthesized at lower temperature (800 $^o$C) have high $j_c$ in high magnetic fields. But in principle high SC performance can be attained for the materials without additions.

As a rule, the materials high-pressure sintered (*ex-situ*) from a $MgB_2$ powder, (3HP, for example, Fig. 1), showed a somewhat lower $j_c$ than those synthesized (*in-situ*) from Mg and B. That seems is due to the formation of larger $MgB_2$ clusters and less homogeneous distribution of oxygen-enriched or other inclusions, which are located around clusters and can reduce connectivity in the material or can prevent from creating the homogeneous pinning (Figs. 6a, b, c). However, at present there is no certain explanation of this phenomenon.

Materials produced under lower pressures using hot pressing (30 MP) and spark plasma sintering (50 MPa) techniques, (6Hot-P and 5SPS, respectively in Fig. 1), demonstrated lower magnetically estimated critical current densities and material's density (Table 1, Fig. 7) than those produced at high pressure, but their $j_c$ are still rather high for the effective application. Besides, lower pressure makes the manufacture of larger blocks and parts of easier.

The material of rings manufactured by HIP, 7 HIP, or pressureless synthesis, 8PL (under 0. 1 MPa of argon) exhibited the least $j_c$ and density (Fig. 1, Table 1), but in general their characteristics are high enough for the application as well.

3.2 Structural features of the materials prepared at high pressure and hot pressing.

Figure 8 presents structures of high-pressure-synthesized (*in situ*) materials aimed to show their density (secondary electron image - SEI) and distribution of oxygen and higher borides inclusions (backscattering electron image - BEI). In backscattering electron regime or at compositional contrast images the brightness increases with increasing mean atomic number, $Z'$, of the irradiated region, where $Z'=\Sigma c_i Z_i$ and $c_i$ is the weight fraction of element $Z_i$. The structure of materials synthesized under high pressure can be near theoretically dense, see, for example, Fig. 8 b. In previous publications [7, 11] we discussed the dependence of critical current density on the amount, distribution, and size of "black" inclusions of higher borides, the stoichiometry of which in high-pressure (2 GPa) synthesized and sintered $MgB_2$ materials is usually near $MgB_{12}$ (Fig. 8) and in materials obtained by hot-pressing (30 MPa) is near $MgB_7$ (Fig. 7). Despite the fact that higher borides are present in the structure of $MgB_2$-based materials in large quantities, they can not be practically revealed in X-ray patterns. This can be explained by poor diffracted signals because of the low X-ray atomic scattering factor of boron [6]; besides, inclusions of higher borides are dispersed in $MgB_2$. The etalon X-ray pattern of $MgB_{12}$ is absent in the database and the literature data are contradictive (some authors mentioned the orthorhombic structure of $MgB_{12}$ [12] and the other ones pointed out that it was rhombohedral or hexagonal [13]). Our numerous observations showed that in the structures of high pressure synthesized $MgB_2$-based materials with higher critical current densities usually higher amount of fine dispersed $MgB_{12}$ inclusions are present. Figures 6d and 8m demonstrate the absence of oxygen-enriched inclusions in the $MgB_2$ structure around grains of higher borides. Up to now we have thought that inclusions of higher borides and $MgB_{12}$, in particular, can play a role of pinning centers in $MgB_2$ but it may be that in addition, they can affect the oxygen distribution in the $MgB_2$ matrix and assist the formation of oxygen-enriched Mg-B-O inclusions, which can influence pinning as well.

Our attempts to find correlations between the total oxygen content in the $MgB_2$ materials matrix, its amount in starting boron or magnesium diboride and critical current density were failed [14]. But it seems that more important for material's superconducting properties is the oxygen distribution mode. By comparing the dependences of $j_c$ (Fig. 2), materials structures (Fig. 8) and results of SEM microprobe analysis it is possible to resume the following. The matrix of all HP-synthesized samples contained Mg and B in near-$MgB_2$ stoichiometry and some amount of oxygen (5-10 wt.% or sometimes even more). Oxygen in $MgB_{12}$ inclusions was practically absent (the fact that we found oxygen in a small amount of 0.2-1.2 wt.% in these inclusions may be the result of "touching" by electron beam the area around the $MgB_{12}$ inclusion). Our structural observations have shown that at a synthesis temperature of about 800 $^o$C oxygen is more homogeneously distributed in the matrix material (Figs. 8d, e), but in the material synthesized at 1050 $^o$C the distribution of oxygen in the matrix is less homogeneous: the oxygen-enriched small areas form (more bright areas in Figs. 8 a, c) and with addition of Ta or Ti such areas are transformed into well delineated dispersed Mg-B-O inclusions (light or white inclusions in Figs. 8f-h). In the materials with SiC additions such Mg-B-O inclusions are observed as well (Fig. 8k, l), they are well seen in places where SiC inclusions are absent. The values of $j_c$ correlate well with such a so-called segregation of oxygen in $MgB_2$-based structures: usually the higher segregation corresponds to higher $j_c$. The matrix of material with these oxygen-enriched inclusions produced at 1050 $^o$C (from Mg and B with 10 wt.% of Ti additions) contains less oxygen (1.5 – 5 %) than that (8 %) in low-temperature produced samples at 800 $^o$C. The results of SEM microprobe study and SEM images (Fig. 9) give proofs that white inclusions formed consist of Mg, B and O and that amount of oxygen in them are higher than that in the surrounding $MgB_2$ matrix.

As seen from the results shown in Fig. 2 and by other our mutual observations, different types of amorphous boron as well as magnesium diboride produced even by the same company after sintering or synthesis under the same high pressure–high temperature conditions demonstrate rather different superconducting characteristics (for example, dependences of $j_c$ on magnetic field), while the reproducibility of these characteristics for each type of boron or magnesium diboride is very high. The difference in oxygen content

(0.66–3.7 %) and materials grain sizes (0.8–9.6 μm) between the initial powders were not the dominant reasons. So, possibly some other admixture in the initial boron or magnesium diboride can affect the materials behavior (with high probability it can be carbon and/or hydrogen), for example, the formation of magnesium hydride [15] or carbon incorporation in the material structure and into magnesium diboride lattice, in particular [10].

The structure of the material with high $H_{c2}$ (4HP in Fig. 1) is very fine and complicated for the investigation (Fig. 4). The dark grey matrix is of near $MgB_4$ composition, black inclusions with near the $MgB_{20}$ stoichiometry are observed, the stoichiometry of the brightest white inclusions or spots are of near MgO composition. Less bright inclusions or spots contained Mg, B and O, for example, the average composition of the area inside square "A" had stoichiometry close to $MgB_{3.1}O_{0.3}$ and the material matrix contains 7 wt% of oxygen. Reflexes marked "x" at X-ray pattern may be assigned to higher borides. Despite the fact that reflex at $2\Theta = 26.8\ °$ may be attributed to hexagonal BN the absence of nitrogen in the material structure and the increase of reflex intensity with the increase of the amount of higher borides and even with the amount of boron in higher boride gave us grounds to associate its appearance with higher borides.

For our observation the improvement of materials $j_c$ in the case of SiC adding occurred when there was no notable interaction between SiC and $MgB_2$ (when the X-ray pattern contains reflexes of SiC, but no reflexes of $Mg_2Si$ and when reflexes of $MgB_2$ are not shifted from the etalon positions). Additions of SiC also seem to promote the oxygen segregation (Figs. 8j-l), besides, SiC grains may produce effective pinning by themselves.

The evolution of the material structure synthesized or sintered under high pressure (2 GPa) allows us to propose the following correlations between the structural features and materials superconductive characteristics. Matrix contains oxygen, the ratio between B and Mg in matrix is near to $MgB_2$. Possibly oxygen solved in $MgB_2$, lattice parameters of magnesium diboride synthesized (ex-situ) or sintered (in situ) in the 700–1100 $^{o}$C temperature range varied as follows: $a$ = 0.30747 – 0.30822 nm and $c$ = 0.35174 – 0.35212 nm and the MgB$_2$ reflexes practically did not shift from their etalon positions. Only small amount of MgO is usually present in the materials and its content does not depend on the temperature- pressure conditions. At low temperature (700–800 $^{o}$C) grains of higher borides are forming (mainly of the MgB$_{12}$ stoichiometry, while grains with composition near MgB$_{16-17}$ or MgB$_{20}$ can also be present), after cooling one can observe dispersed grains of higher borides in the MgB$_2$ matrix. Larger amount of MgB$_{12}$ and smaller MgB$_{12}$ grains are usually observed in materials with higher $j_c$ (especially in high magnetic fields). Oxygen is practically absent in the structure of higher borides and during crystallization it seems that higher boride grains even promote the oxygen "pushing away" from the MgB$_2$ matrix surrounding the inclusions and thus cleaned the MgB$_2$ grain boundaries from oxygen (see, for example, Fig. 6d and 8m). Additions of Ti, Ta or Zr stimulate the formation of higher borides and absorb admixture hydrogen at low manufacturing temperatures, prevent harmful (for material $j_c$) MgH$_2$ from being formed and thus, affect the oxygen distribution (segregation) stimulating the formation of Mg-B-O inclusions, which in turn can act as pinning centers in the material.

At higher temperatures (900–1050 $^{o}$C) even under high pressure (2 -5 GPa) admixing hydrogen usually "leave" the material structure. As the manufacturing temperature increases, the amount of higher borides decreases but oxygen is redistributed into separate dispersed Mg-B-O areas or inclusions (especially, when Ti or Ta is added), which can promote pinning. High manufacturing temperatures, as a rule, result in an increase of $j_c$ in low and medium magnetic fields.

Oxygen does not usually solve in the structure of higher borides [7] but it can be incorporated in the MgB$_2$ structure even without its essential disturbance [16]. According to

[13], the reaction in Mg-B system starts from the $MgB_{12}$ formation and then $MgB_2$ forms, so grains of $MgB_{12}$ are possibly, centers of the $MgB_2$ phase crystallization. As it was shown in [17] Ti, Ta, and Zr at low synthesis temperatures can react with hydrogen to form hydrides ($Ta_2H$, $TiH_{1.924}$, $ZrH_2$), thus preventing the $MgH_2$ forming and promoting the liberation of Mg, which may react with boron. Because of this larger amount of dispersed $MgB_{12}$ inclusions or other higher borides can appear. As the synthesis temperature increases, Mg intensively diffuses toward grains of $MgB_{12}$ or higher borides and react with them, so, they become smaller due to the formation of $MgB_2$ at their periphery and because oxygen is absent in the structures of higher borides, the forming $MgB_2$ matrix phase contains much less oxygen and, may be because of this, it seems that oxygen or oxygen-enriched inclusions are "pushed out" from the area around the higher boride inclusion. So, on the one hand, inclusions of higher borides (or grain boundaries between $MgB_2$ matrix and higher boride inclusions) can increase pinning, and on the other, higher borides can promote the "cleaning" of $MgB_2$ intragrain boundaries from oxygen, thus improving connectivity and superconducting characteristics, $j_c$, in particular.

3.4 Raman spectroscopy study of Mg-B -based materials.

The presence of higher borides may influence the critical current density of magnesium diboride-based materials. However, there exists the vagueness concerning appearance, composition and structure of higher boride phases ($MgB_4$, $MgB_6$ or $MgB_7$, $MgB_{12}$, $MgB_{16}$ or $MgB_{17}$ and $MgB_{20}$) in the literature.

The author of [12] reported the synthesis of an orthorhombic $MgB_{12}$ single crystal, space group Pnma with $a = 1.6632(3)$ ° nm, $b = 1.7803(4)$ nm and $c = 1.0396(2)$ nm from the elemental magnesium and boron in a Mg/Cu melt at 1600 °C (Cu, Mg and B mixed in the molar ratio of 5:3:2) and claimed that according to the preliminary results, the single crystalline samples were non-superconducting down to 2K. According to [13], the structure of the synthesized $MgB_{12}$ and $MgB_{20}$ materials was rhombohedral or trigonal with lattice parameters $a$ = 1.1014(7) nm, $c$ = 2.4170(2) nm) and $a$ = 1.09830(4) nm , $c$ = 24.1561(2) nm, respectively. $MgB_7$, crystallizes in the space group Imam with $a$ = 0.5597(3) nm, $b$ = 8.125(3) nm , and $c$ = 10.480(5) nm.

We synthesized materials with superconducting characteristics from the mixtures of Mg and B taken in the 1:4, 1:6, 1:8, 1:10, 1:12, 1:17, 1:20 ratios exposed to 2 GPa and 800 or 1200 °C for 1 h. In the case when the stoichiometry of the resulting material matrix was near $MgB_{12}$ (confirmed by the SEM and TEM observations) the critical current density was rather high (up to 55 kA/cm$^2$ in zero field and $H_{irr}$=7.2 T at 10 K and up to 45 kA/cm$^2$ in zero field and $H_{irr}$=4,7 T at 20 K), and the materials had sharp SC transition at about 37 K (determined from the imagined and real parts of the AC susceptibility curves) [11]. The materials with the highest $j_c$ and amount of superconducting phase (among those prepared from mixtures with higher boron content than in case of Mg:B =1:2) were materials synthesized from mixtures with the Mg:B ratios of 1:8 [11] and 1:20 (Figs. 10b, d, f, h) at 2 GPa, 1200 °C for 1 h. But when the Mg:B =1:8 mixture was exposed to 5.5 GPa, 1200 °C for 24 h we got essentially nonsuperconducting material (Fig.10g), which structure according to X-ray consisted mainly of the $MgB_7$ phase and rather big amount of MgO (Fig. 10a). However the SEM EDX study revealed that the matrix of this material was close to $MgB_{12}$ in composition (Fig. 10e). Despite the fact that both matrixes had near $MgB_{12}$ stoichiometry according to the SEM study (Figs. 10e, f), their superconducting properties (10g, h) and Raman spectra (10c, d) are very different. X-ray patterns of these samples are rather different as well. It should be noted that very often a large amount of higher borides, which are present in the materials, cannot be revealed by X-ray diffraction because of the dominating reflexes of $MgB_2$ phase despite its fraction in the samples is very small. It seems that in contrary to X-ray, Raman spectroscopy may be of great help in distinguishing between phases of different stoichiometry in such cases.

3.5 Mechanical characteristics of high pressure prepared Mg-B-based materials.

The hardness, fracture toughness and Young modulus of high-pressure manufactured materials were estimated by indentation methods. Despite the fact that an indentation load is included into the formulas for determining the mechanical properties, in the case of ceramic materials it is important to mention the indentation load because of the elastic aftereffect and changing of the indent shape after removing the indenter. The elastic aftereffect results in an increase of the estimated values of microhardness and fracture toughness as the indentation load decreases. The hardness or microhardness (estimated under a comparatively high load) is an integrated characteristic and gives a notion about the mechanical properties of a material as a whole, because the area of the indent is rather large and many grains appear under the indent. Using the nanohardness one can characterize mechanical properties of separate phases, which are present in the material if the size of their grains or the area occupied by the phase is enough for performing the study.

The microhardness of the material sintered at high pressure (2 GPa) and high temperature (800 $^{o}$C), determined using a Vickers indenter ($H_v$) under a load of 4,9 N was: 12.65±1.39 GPa (for the material without additions), 13.08±1.07 GPa for 10% Ta addition and 12.1 ±0.08 GPa for 2% Ti addition. The Vickers hardness under a 148.8 N load was $H_v$=10.12±0.2 GPa. Because of the absence of cracks under a load of 4.9 N, the fracture toughness was estimated at a load of 148.8 N: $K_{1C}$=4.4± 0.04 MPa·m$^{0.5}$ for the material without additions and $K_{1C}$=7.6± 2.0 MPa·m$^{0.5}$ for the material with addition of 10% Ta.

The mechanical properties (Berkovich nanohardness and Young modulus) of $MgB_{12}$ inclusions (of about 10 μm in size) in the $MgB_2$ matrix of high-pressure sintered material were estimated using a Nano Indenter II nanohardness tester. The nanohardness of inclusions with near $MgB_{12}$ stoichiometry at a 60 mN-load is 35.6±0.9 GPa, which is higher than that of sapphire (31.1±2.0 GPa), while the nanohardness of the "matrix" under the same load was 17.4±1.1 GPa only. The Young modulus is 213±18 GPa for the "matrix", 385±14 GPa for the inclusion and 416± 22 GPa for sapphire.

The Vickers microhardness ($H_v$) of the material with near $MgB_{12}$ composition, prepared at 2 GPa and 1200 °C for 1 h was twice as high as that of $MgB_2$ (25±1.1 GPa and 12.1±0.8 GPa, respectively, at a load of 4.9 N).

The addition of Ti or Ta makes $MgB_2$-based materials high-pressure produced at low temperatures (700–800 °C) denser and more mechanically stable due to the possibility to absorb admixture hydrogen and thus, prevent pores and crack from being formed (especially in large blocks).

3.5 Applicability of $MgB_2$-based materials for current limitation.

Figure 11 demonstrates the results of testing the ring cut from hot pressure (Hot-P) synthesized block. The contactless transformer method is suitable for samples in the form of hollow superconducting cylinders and rings and allows one to achieve high currents flowing in the superconductor using the usual laboratory equipment. A ring forms the secondary coil of a transformer, in which the primary coil is connected with a circuit of an AC source. To increase the coupling between the coils, the ring and primary coil are centered on a ferromagnetic core. This method may be related to AC loss measurements in samples with a transport current. The losses can be calculated from the primary coil current and voltage waveforms. The analysis of these waveforms also allows studying the peculiarities of the magnetic field penetration and quenching [8-9].

The dimensions of the tested rings were the following for ring 1: outer diameter – 21.3 mm, height - 14.1 mm, wall thickness – 3.5 mm and for ring 2: outer diameter – 45 mm, height - 11.6 mm, wall thickness – 3.3 mm (Fig. 11e). Ring 1 was tested in the temperature range from 4.2 to 40 K; ring 2 – at 4.2 K (liquid helium). The behavior of the rings is similar and qualitatively independent of temperature. As an example the results for ring 2 are presented in Figs. 11. The quenching current determined from response of the transformer devices, in which the secondary winding was fabricated as a superconducting ring, to a current pulse. The quenching current is determined as a current corresponding to jump on the voltage trace and is estimated at 24000 A. The critical current density is 63200 A/cm$^2$. This value is about an order of magnitude lower than the critical value obtained from magnetization experiment ($6 \cdot 10^5$ A/cm$^2$). This can be explained by the granular structure of the superconductor, when the critical current density measured from magnetization is mainly determined by this density in granules but the transport critical current is determined by the properties of intergranular area. At a lower voltage of source (125 V) (Fig. 11d) quenching is observed after several periods of a current, during which the cylinder is heated. These losses till quenching are about 17 J (Fig. 11b). Power of the losses is about 200 W. The VSM tests have shown that SC characteristics (dependences of $j_c$ on H) of the HP-synthesized MgB$_2$-based material remained practically unchanged after 6 years of keeping in air at room temperature.

Cooling of the high-temperature superconducting cylinders from BSCCO takes a prolonged time due to their low thermal conductivity (~ 1 W/K m) and after fault current limitation superconducting properties do not recover for about 1 s as it is required for application of these FCLs in power systems. The critical current of the BSCCO ring with a diameter of 20 cm, thickness of 0.5 cm, and height of 10 cm is about 3.500 A at 77 K. At the same time the critical current of the MgB$_2$ ring with a diameter of 4.5 cm, thickness of 0.33 cm and height of 1.2 cm is about 2.4 10$^4$ A at 4–10 K.

The thermal conductivity of HP–synthesized MgB$_2$–based materials is much higher (53 ±2 W/(m×K) at 300 K, Table 2) than of BSCCO, which is considered to be promising for application in the transformer-type fault current limiters. The higher thermal conductivity and much more homogeneity of the superconducting properties of MgB$_2$ synthesized under pressure (Fig. 5, curves 1–3) should prevent the creation of overheated domains in the condition when BSCCO can be destroyed.

CONCLUSIONS

High pressure (2 GPa) synthesized or sintered materials under optimal conditions exhibited the higher critical current densities (estimated magnetically) and the lower porosity than hot-pressed (30 MPa) or SPS-manufactured (50 MPa) ones. Synthesized *in situ* MgB$_2$-based materials usually demonstrate a somewhat higher $j_c$ than sintered ex situ. But for practical application all these materials have rather high superconducting and mechanical characteristics. It is surprising that a low-temperature (600 °C) high-pressure synthesized material without additions having rather high porosity (around 17 %) demonstrated at 20 K extremely high H$_{c2}$ and H$_{irr}$ (the highest ever observed, even higher than that of C-doped materials). By varying synthesis or sintering conditions as well as the amount and type of additives (Ti, Ta, SiC) we can essentially affect critical current density of MgB$_2$-based materials, may be it is because of the variation of the oxygen distribution in the materials structure and the formation of dispersed Mg-B-O regions or inclusions with high concentration of oxygen. Besides, it is possible to promote the higher borides formation in MgB$_2$-based material matrix, influence their size and distribution, which in turn can affect $j_c$.

But we cannot exclude that the higher borides du0ring their formation influence the oxygen distribution in material matrix as well. The higher synthesis temperatures promote oxygen redistribution and segregation. In the case of SiC additions the increase of $j_c$ was observed when there was no interaction between $MgB_2$ and SiC notable by the X-ray analysis and it seems that inclusion of SiC together with segregate oxygen-enriched inclusions can affect pinning and may promote increasing of $j_c$. Despite the possibility to influence SC characteristics using additions or varying manufacturing temperature our numerous experiments have shown that the determinative role in attaining high superconductive characteristics of $MgB_2$-based materials is played by the quality of initial boron or magnesium diboride. Up to now it is not clear what characteristic or impurity played the main role in the tested initial materials (several types of amorphous boron containing 0.66–3.7 % oxygen and with average grain sizes of 0.8–4 μm and magnesium diboride with 0.8–3.5% oxygen and 0.8–9.6 μm grains). It can be admixture of carbon or hydrogen, for example.

It has been shown that higher borides can be revealed in $MgB_2$ using Raman spectroscopy. The Raman spectrum of that material that according to X-ray had $MgB_7$ composition (with some impurity MgO) has been obtained. But in contradiction with X-ray data the SEM EDX study shows that the material had near $MgB_{12}$ stoichiometry.

The tests of rings from the $MgB_2$ demonstrated perspectives of their application in transformer type fault current limiters. The high values of $H_{c2}$ attained make it possible the use of bulk $MgB_2$ in high magnetic fields, for manufacturing of magnets, etc.

References

1. C. Buzea and T. Yamashita, Supercond. Sci. Technol. 14, 115 (2001).
2. M. Eisterer, Supercond. Sci. Technol. 20, 47 (2007).

3. E. W. Collings, M. D. Sumption, M. Bhatia, M. A. Susner and S. D. Bohnenstiehl, Supercond. Sci. Technol. 21, 103001 (2008).

4. M. Eisterer, M. Zehetmayer, S. Tönies, H. W. Weber, M. Kambara, N. Hari Babu, D. A. Cardwell and L. R. Greenwood, Supercond. Sci. Technol. 15, L9 (2002).

5. R. Flükiger, M. S. A. Hossain, C. Senatore, *cond-mat /*0901.4546 (for SUST) (2009).

6. B. Birajdar, N. Peranio and O. Eibl, Supercond. Sci. Technol. 21, 073001 (2008).

7. T. A. Prikhna, W. Gawalek, Ya. M. Savchuk, T. Habisreuther, M. Wendt, N. V. Sergienko, V. E. Moshchil, P. Nagorny, Ch. Schmidt, J. Dellith, U. Dittrich, D. Litzkendorf, V. S. Melnikov and V. B. Sverdun, Supercond. Sci. Technol. 20, S257, (2007).

8. V. Meerovich, V. Sokolovsky, G. Grader, G.Shter, Appl. Supercond. 2, 123 (1994).

9. V. Meerovich, V. Sokolovsky, J. Bock, 5, 22 (1995).

10. M. Mudgel, L. S. S.Chandra, V. Ganesan, G. L. Bhalla, H. Kishan, V. P. S. Awana, J. Appl. Phys., 106, 033904 (2009)

11. T. A. Prikhna, W. Gawalek, Ya. M. Savchuk, A. V. Kozyrev, M. Wendt, V. S. Melnikov, V. Z. Turkevich, N. V. Sergienko, V. E. Moshchil, J. Dellith, Ch. Shmidt, S. N. Dub, T. Habisreuther, D. Litzkendorf, P. A. Nagorny, V. B. Sverdun, H. W. Weber, M. Eisterer, J. Noudem and U. Dittrich, IEEE Transactions on Applied Superconductivity 19, 2780 (2009).

12. Adasch V. 2005 Dissertation am der Fakultat fur Biologie, Chemie und Geowissenschaften der Universitat Bayreuth «Synthese, Charakterisierung und Materialeigenschaften borreicher Boride und Boridcarbide des Magnesiums und Aluminiums» unter der Anleitung von Prof. Dr. Harald Hilleb

13. R. Schmitt, J. Glaser, T. Wenzel, K. G. Nickel, H.-Ju. Meyer, Physica C, 436, 8 (2006)

14. T. Prikhna, W. Gawalek, Y. Savchuk, N. Sergienko, M. Wendt, T. Habisreuther, V. Moshchil, A. Mamalis, J. Noudem, X. Chaud, V. Turkevich, P. Nagorny, A. Kozyrev, J.

Dellith, C. Shmidt, D. Litzkendorf, U. Dittrich and S. Dub, J. Optoelectron.Adv. Mater. 10, 1017 (2008).

15. T. A. Prikhna, W. Gawalek, Ya. M. Savchuk, V. E. Moshchil, N. V. Sergienko, T. Habisreuther, M. Wendt, R. Hergt, Ch. Schmidt, J. Dellith, V. S. Melnikov, A. Assmann, D. Litzkendorf, P. A. Nagorny, *Physica C* 402, 223 (2004).

16. X. Z. Liao, A. C. Serquis, Y. T. Zhu, J. Y. Huang, L. Civale, D. E. Peterson, F. M. Mueller and H. Xu, Journal of Applied Physics 93, 6208(2003).

17. T. Prikhna, N. Novikov, Ya. Savchuk, W. Gawalek, N. Sergienko, V. Moshchil, M. Wendt, V. Melnikov, S. Dub, T. Habisreuther, Ch. Schmidt, J. Dellith, P. Nagorny, Innovative Superhard Materials and Sustainable Coatings for Advanced Manufacturing. Edited by Jay Lee, Nikolay Novikov. NATO Science Series. II. Mathematics, Physics and Chemistry 200, 81 (2005).

**Figure captions**

**Figure 1.** Dependences of critical current density (magnetic), $j_c$, on magnetic field, $\mu_o H$, for MgB$_2$-based materials at 20 K (a) and 30 K (b):
1) HP – high-pressure-synthesized at 2 GPa, 1050 °C, 1 h from Mg (I) and B (I) taken into the MgB$_2$ stoichiometry with a 10% SiC addition;
2) HP – high-pressure-synthesized at 2 GPa, 1050 °C, 1 h from Mg (I) and B (I) taken into the MgB$_2$ stoichiometry;
3) HP – high-pressure-sintered at 2 GPa, 1050 °C, 1 h from MgB$_2$ ;
4) HP – high-pressure-synthesized at 2 GPa, 600 °C, 1 h from Mg (II) and B (II) taken into the MgB$_2$ stoichiometry;
5) SPS – spark-plasma-synthesized under 50 MPa, at 600 °C for 0.3 h and at 1050 °C for 0.5 h from Mg(I) and B(III) taken into the MgB$_2$ stoichiometry;
6) HOT-P – synthesized by hot pressing at 30 MPa, 900 °C, 1 h from Mg(I) and B(III) taken into the MgB$_2$ stoichiometry with a 10% Ta addition;
7) HIP – synthesized under high isostatic (gas) pressure at 0.1 GPa, 900 °C, 1 h from a precompacted into a ring by broaching mixture of Mg(I) and B(III) taken into the MgB$_2$ stoichiometry with a 10% Ti addition;
8) PL – pressureless sintering at 0.1 MPa (1 atm of Ar), 800 °C, 2 h from a precompacted into a ring by broaching mixture of Mg(I) and B(III) taken into the MgB$_2$ stoichiometry with 10 wt.% of Ti added.

**Figure 2.** (a, b) Dependences of critical current density (magnetic), $j_c$, at different temperatures on magnetic field, $\mu_o H$ at 20 K for materials synthesized from Mg and B taken in the MgB$_2$ stoichiometry without and with dopants under a high pressure of 2 GPa for 1 h at $T_s$ = 800 and 1050 °C ;

**Figure 3**. Upper critical field, $H_{c2}$ and the field of irreversibility, $H_{irr}$, as a function of temperature, T, for high-pressure synthesized materials (at 2 GPa for 1 h) from Mg and B before and after irradiation by fast neutron fluence of $10^{22}$ m$^{-2}$ (E > 0.1 MeV):
1. Mg(I):B(IV)=1:2 + 10 wt.% Ti, at 800°C,
2. Mg(I):B(IV)=1:2 + 10 wt.% Zr, at 800°C,
3. Mg(I):B(IV)=1:2 + 10 wt.% Ti, at 800°C after irradiation,
4. Mg(I):B(IV)=1:2 + 10 wt.% Zr, at 800°C after irradiation,
5. Mg(II):B(II)=1:2, at 600° C.

**Figure 4.** (a–c) SEM images and (d) X-ray pattern of the high-pressure synthesized at 2 GPa, 600 °C, 1 h material from Mg(II):B(II)=1:2 (a) SEI-secondary electron and (b) BEI-backscattering electron images of the same place under the same magnification; (c) BEI image: D-admixtures of SiC or CaCO$_3$ trapped in material's pores from polishing. The average composition of the area inside square "A" had near the MgB$_{3.1}$O$_{0.3}$ stoichiometry. Reflexes marked "x" at the X-ray pattern may be assigned to higher borides.

**Figure 5.** Real (m') part of the ac susceptibility (magnetic moment) vs. temperature, T, measured in ac magnetic field with 30 μT amplitude, which varied with a frequency of 33 Hz, of materials HP-synthesized from Mg and amorphous B, taken in the MgB$_2$ stoichiometry:
1 - edge of the sample 63 mm in diameter, Mg(I):B (I and III) = 1:2 +2 wt.% Ti, at 2 GPa, 800 °C, 1h
2 – center of the same sample (see description Fig. 4a)
3 - sample 63 mm in diameter, Mg(I):B(III) = 1:2 at 2 GPa, 950 °C, 1h

4 – sample 9 mm in diameter, Mg(I):B(V) =1:2 + 10 wt. % Ti at 2 GPa, 1050 °C, 1h
5 - sample 9 mm in diameter, Mg(I):B(III)=1:2 + 10 wt. % Ti at 2 GPa, 800 °C, 1h
6 - sample 9 mm in diameter, Mg(I):B(III)=1:2 + 10 wt. % Ti at 2 GPa, 1050 °C, 1h
7 - sample 9 mm in diameter, Mg(I):B(III)=1:2 + 10 wt. % Ta at 2 GPa, 1050 °C, 1h
8 - sample 9 mm in diameter, Mg(II):B(II)=1:2 at 2 GPa, 600 °C, 1h

**Figure 6.**
Structure of high pressure synthesized (at 2 GPa) materials: (a, b) in polarized light using an optical microscope and (c, d) in backscattering electron image regime (BEI) obtained by SEM: (a) of sintered from $MgB_2$ (II) at 2 GPa, 900 °C for 1 h; (b) of synthesized from Mg(I)+B(V) =1:2 at 800 °C for 1 h; (c) of sintered from $MgB_2$ (III) at 1100 °C and, (d) of synthesized from Mg(I)+B(I) =1:2 at 1000 °C for 1 h;

**Figure 7.** Structure (a-c) and X-ray pattern (d) of a hot-pressed (at 30 MPa, 900 °C, 1 h from Mg(I) and B(III) taken into the $MgB_2$ stoichiometry with a 10% Ta addition) plate (58×58×10 mm), the $j_c$ value of which is depicted by curves 6 Hot-P in Fig. 1.

**Figure 8.** Structure of HP-synthesized under 2 GPa for 1 h materials from Mg and B taken in the $MgB_2$ stoichiometry without and with additions:
(a, b, c) from Mg (I)+B(III)= 1:2 at $T_S$=1050 °C: (a, b) the same place under the same magnification in backscattering electron image – BEI (a) and secondary electron image-SEI (b), (c) BEI, higher magnification;
(d, e) from Mg (I)+B(III)= 1:2 at $T_S$=800 °C in BEI at different magnifications;
(f) from Mg (I)+B(III)= 1:2 + 10 wt.% Ta at $T_S$=1050 °C in BEI;
(g, h) from Mg (I)+B(III)= 1:2 + 10 wt.% Ti at $T_S$=1050 °C in BEI at different magnifications;
(i) from Mg (I)+B(III)= 1:2 + 10 wt.% Ti at $T_S$=800 °C in BEI;
(j, k) from Mg (I)+B(III)= 1:2 + 10 wt.% SiC at $T_S$=1050 °C in BEI under different magnifications, Fig. 8l shows a place (A) where SiC inclusions are absent;
(l) from Mg (I)+B(I)= 1:2 + 10 wt.% SiC at $T_S$=1050 °C in BEI, the place where SiC inclusions are absent.

**Figure 9.** (a-d) electron image and analysis of elements distribution (the brighter looks the area, the higher is the amount of the element under study) over the area of a sample synthesized from Mg(I):B(III)= 1:2 + 10 wt.% Ti at 2 GPa, 1050 °C, for 1 h: (a) electron image, (b - d) distribution of boron, oxygen and magnesium, respectively (the same sample is shown in Figs. 8 g, i).

**Figure 10.** (a, c, e, g) samples synthesized from Mg( I):B(III)= 1:8 at 5.5.GPa, 1200 °C, 24 h and (b, d, f, h) samples synthesized from Mg( I):B(III)= 1:20 at 2 GPa, 1200 °C, 1 h:
(a, b) X-ray patterns: reflexes attributed to $MgB_7$ are marked by "1", to MgO by "2", and to $MgB_2$ by "3", the unidentified reflexes are marked by "X";
(c, d) Raman spectra
(e, f) SEM backscattering electron images, D – MgO or SiC from polishing trapped in pores,
(g, h) dependences of critical current density (magnetic), $j_c$, at different temperatures on magnetic field, $\mu_o H$.

**Figure. 11.** Characteristics of the $MgB_2$-based material synthesized at 30 MPa, 800 °C, 2 h from amorphous boron (III) (H.C. Starck, 1.5 % O, 4 μm grains) and magnesium (I) taken in the $MgB_2$ stoichiometry and of the ring cut from this material:

(a) dependences of critical current density (magnetic), $j_c$, on magnetic field, $\mu_o H$;

(b) losses in the ring determined from the experimental data, the initial temperature being 4.2 K, losses are calculated for the case when voltage of the source was 125 V;

(c, d) typical oscilloscope traces of the current (solid lines) and voltage drop across the primary normal-metal winding (dashed lines). Arrows show the quenching current. Voltage of source was 170 V (c) and 125 V (d);

e) the general view of the tested ring: outer diameter – 45 mm, height - 11.6 mm, wall thickness – 3.3 mm.

Table 1 Porosity of MgB$_2$-based materials prepared by different methods

| Under high pressure (2 GPa), HP | | Hot pressing (30 MPa), Hot-P | Spark plasma synthesis, (50 MPa), SPS | High isostatic pressing (0.1 GPa), HIP | Pressure less synthesis (0.1 MPa) after compacting by broaching, PL |
|---|---|---|---|---|---|
| 800 - 1100 °C | 600 °C | | | | |
| 1 – 3.5 % | 17 % | 8 - 28 % | 3 - 20 % | 44% | 47 % |

Table 2
Thermal conductivity of MgB$_2$-based material at 300 K

| High-pressure synthesized at 2 GPa, 800 $^o$C, 1 h from Mg(I):B(III)=1:2 + 10 wt.% Ti | Thermal conductivity λ, W/(m·K) |
|---|---|
| square sample 15×15×2 mm | 53±2 |
| rectangular sample 15×20×2 mm: along 15 mm side along 20 mm side | 45±2 44±2 |

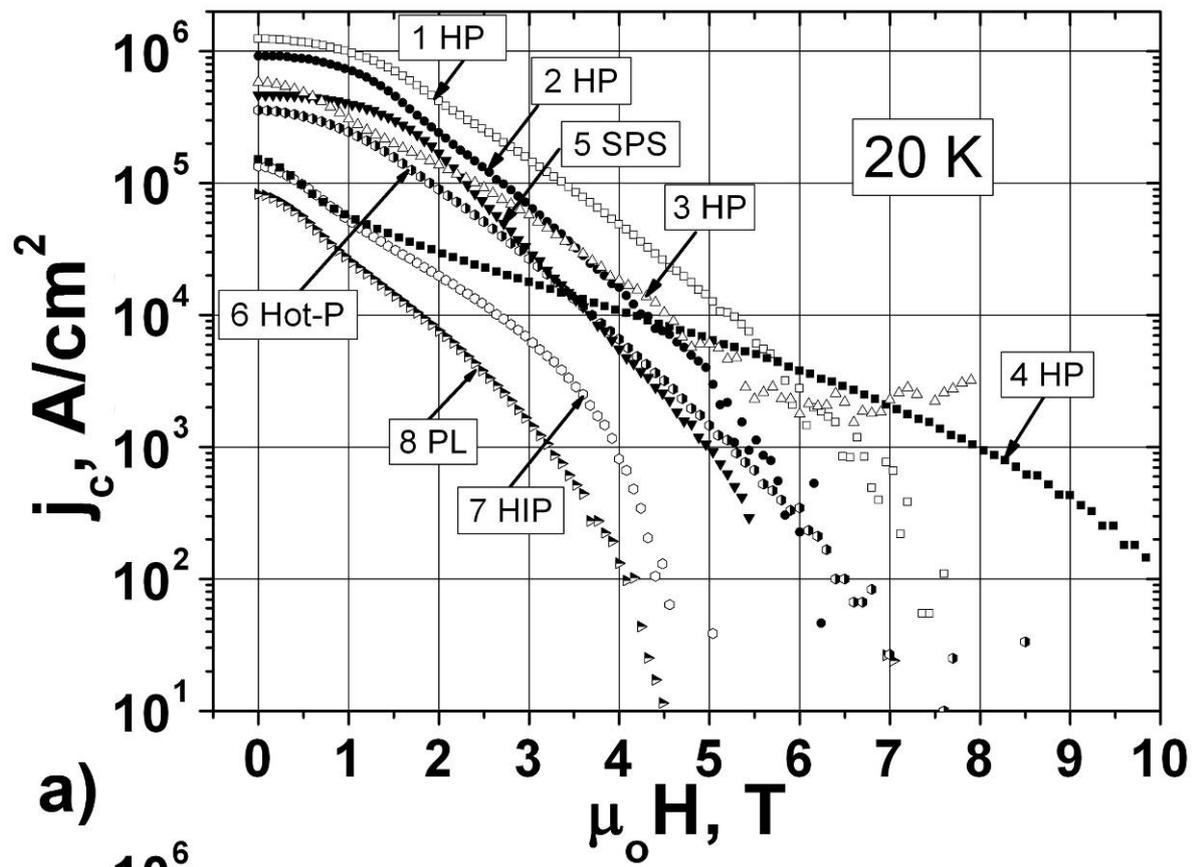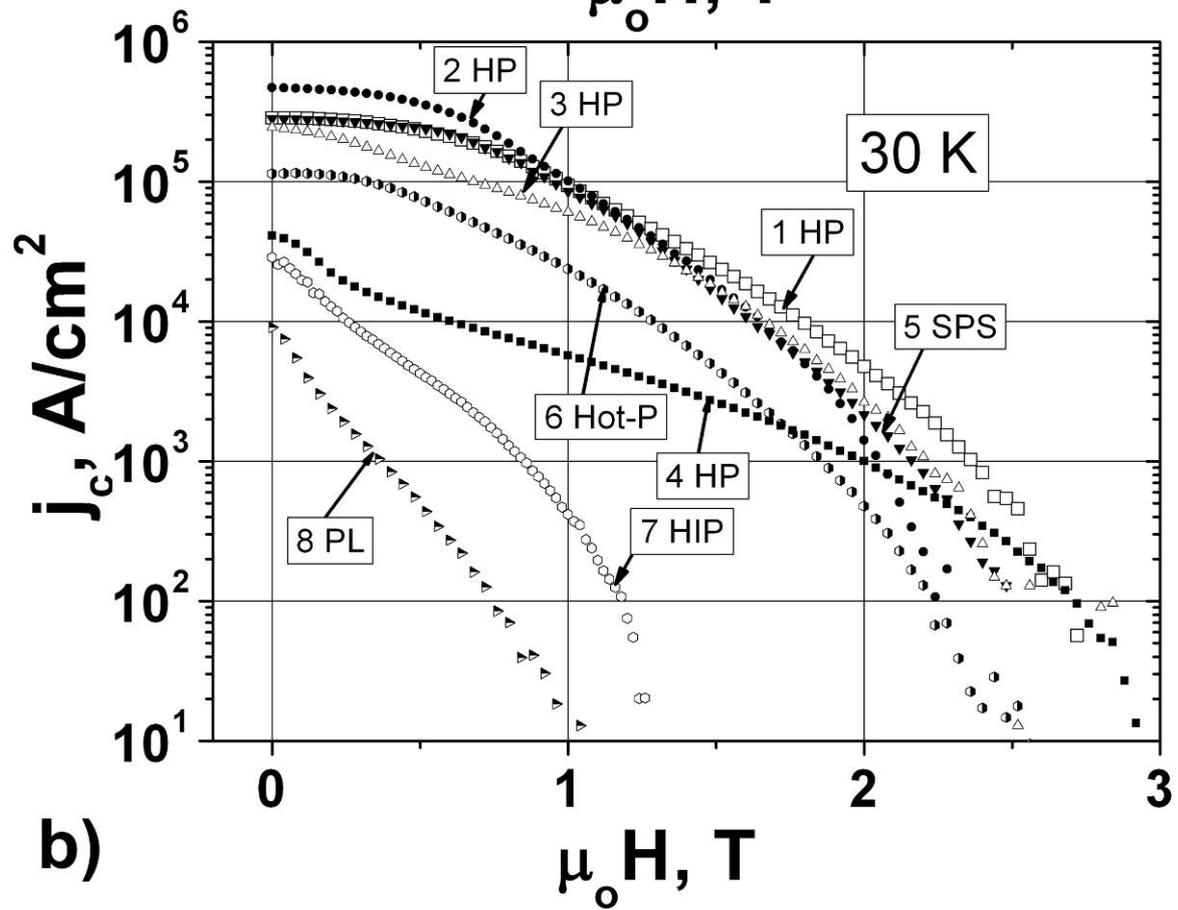

Figure 1

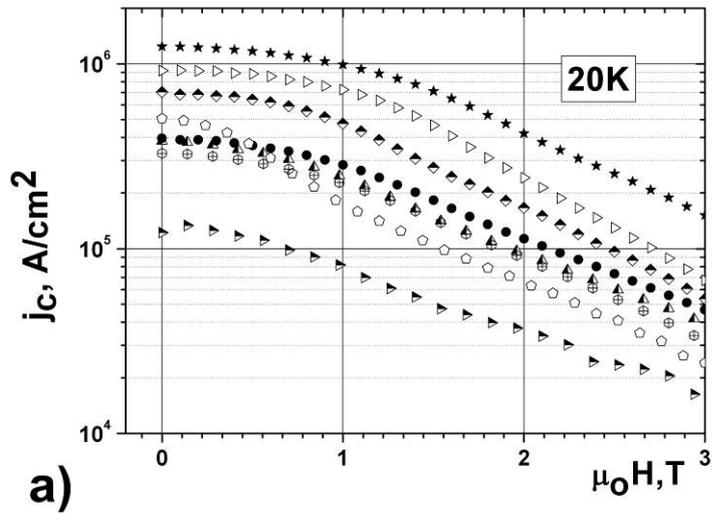
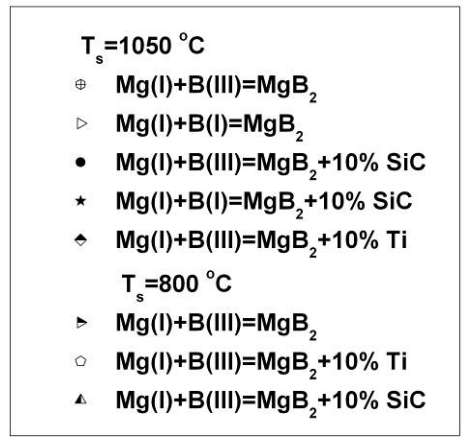
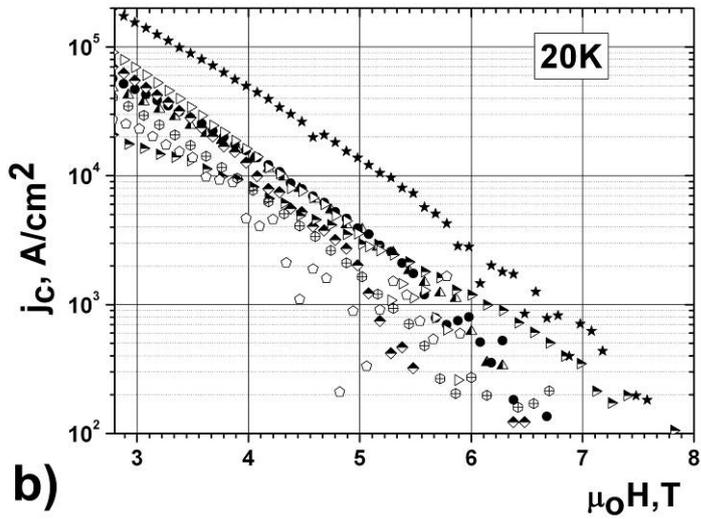
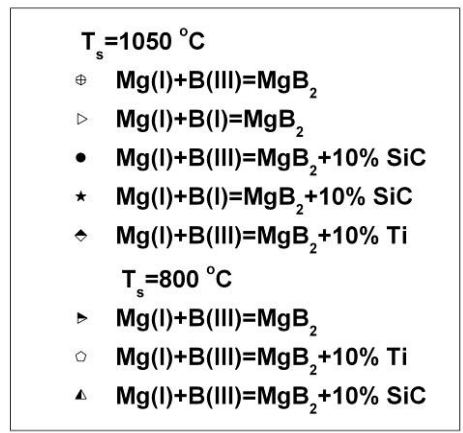

**Figure 2**

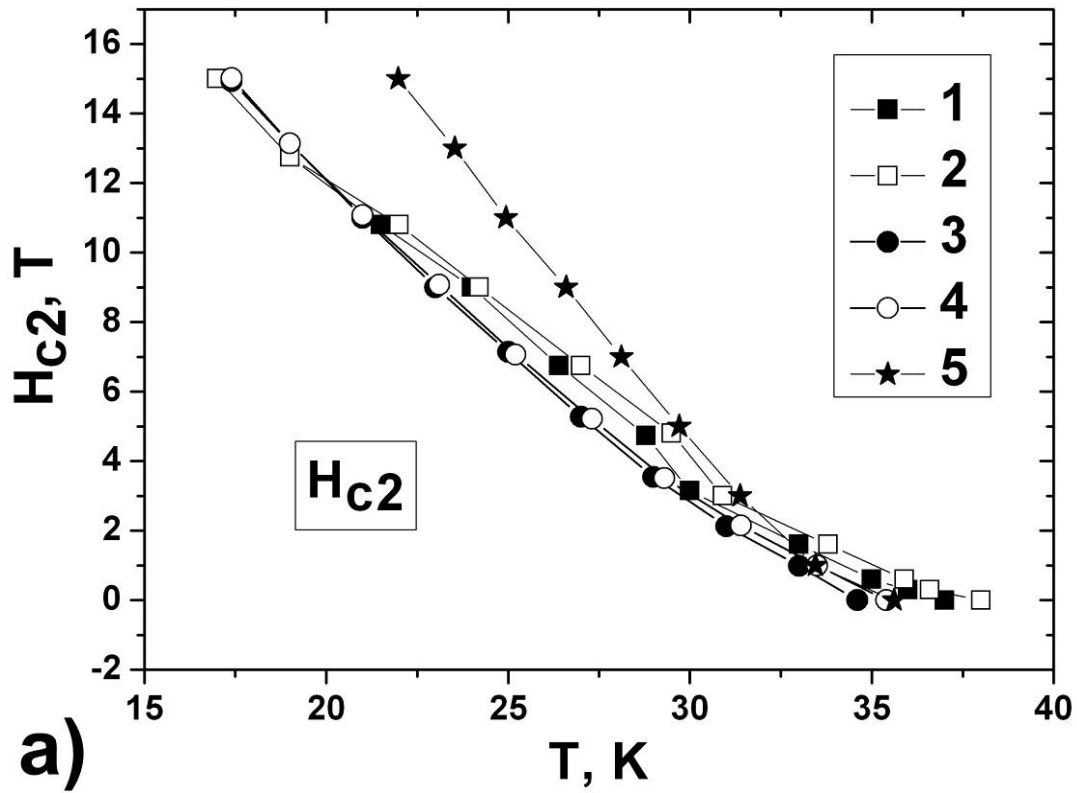

a)

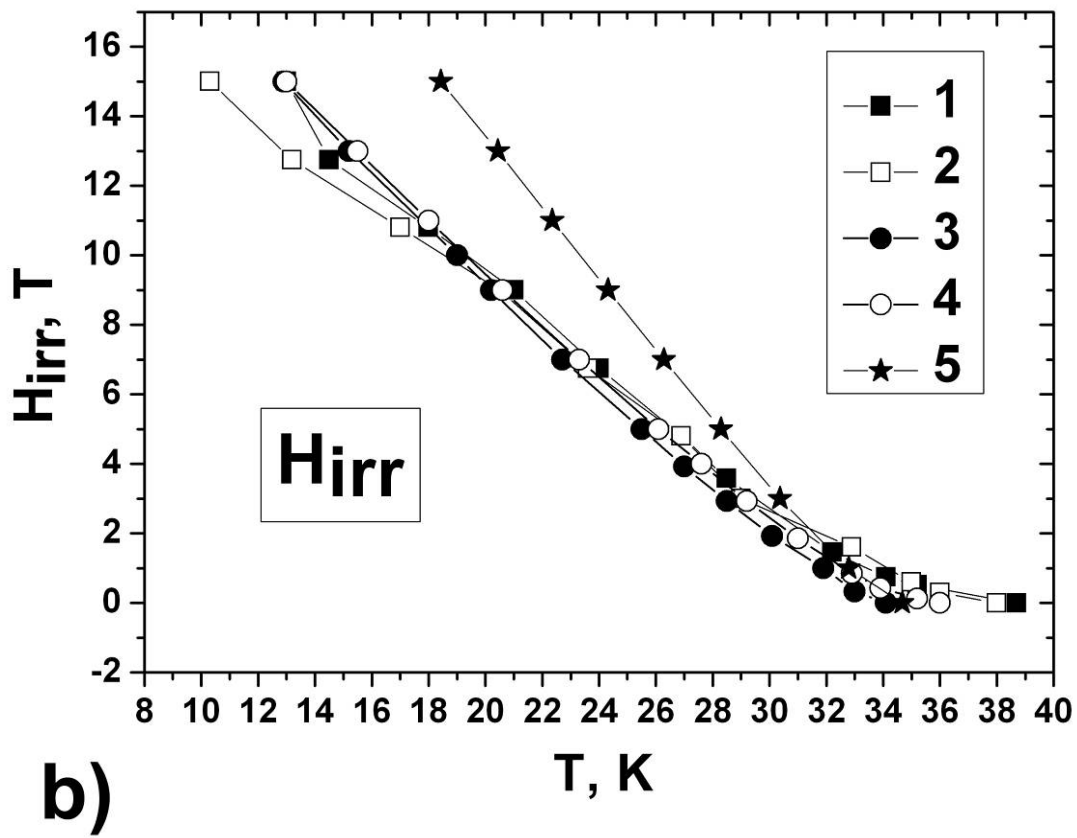

b)

**Figure 3.**

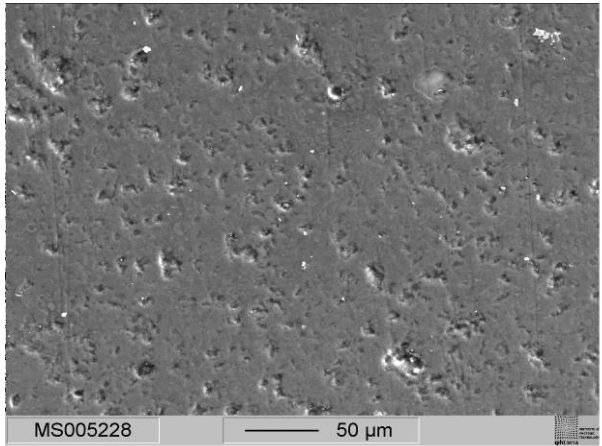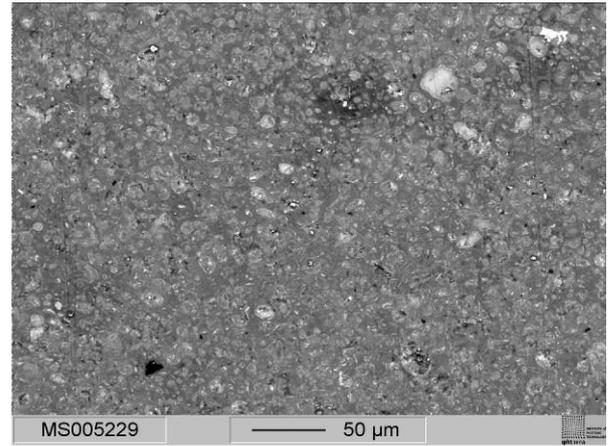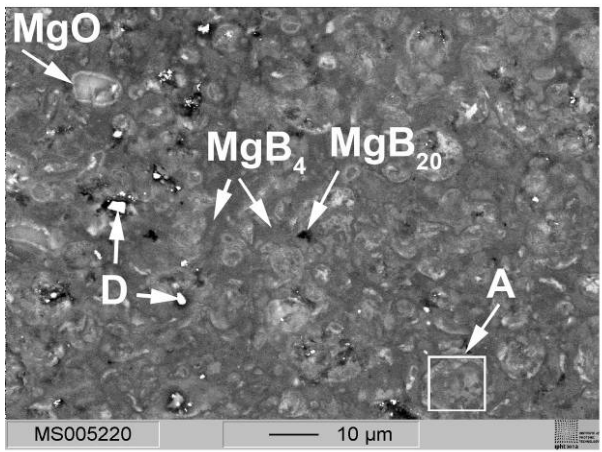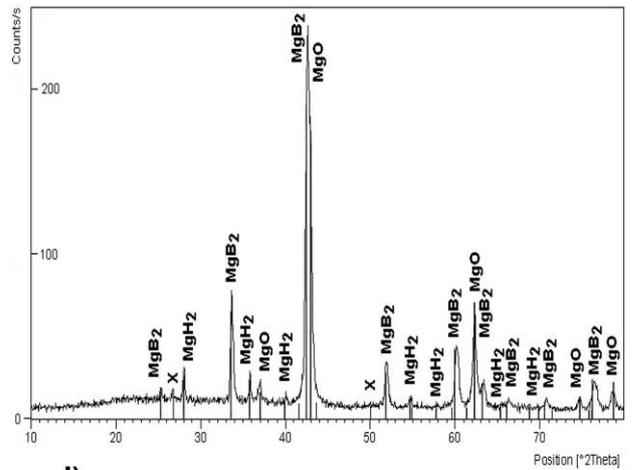

**Figure 4**

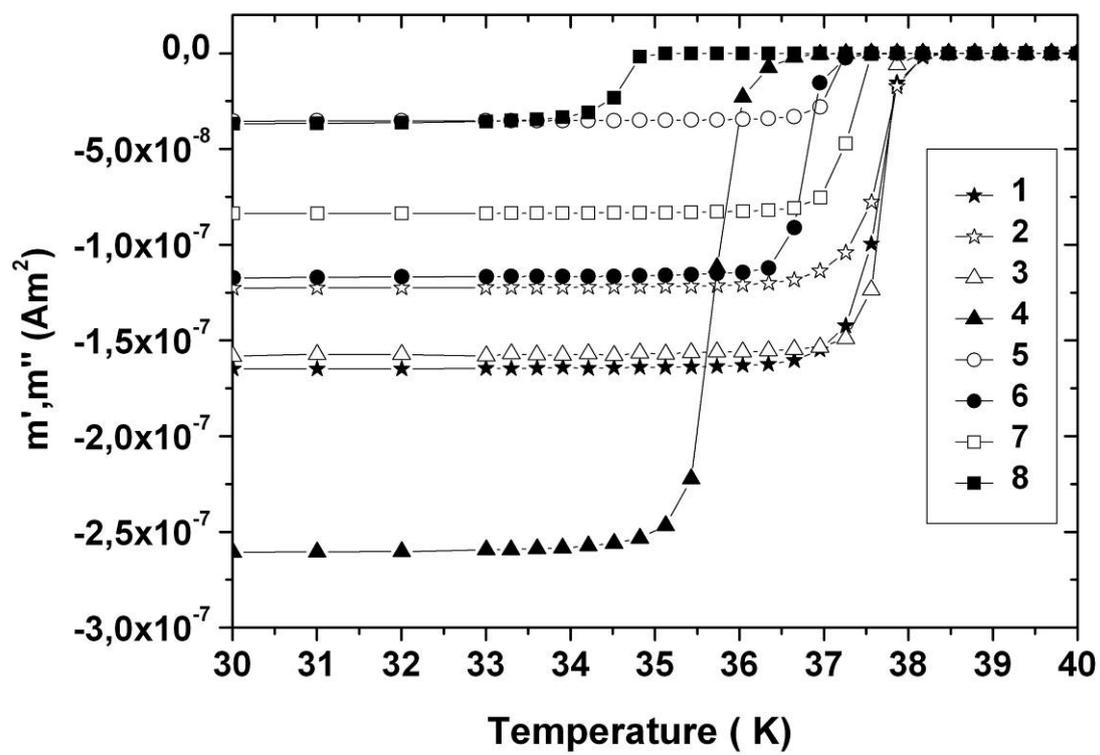

**Fgure 5**

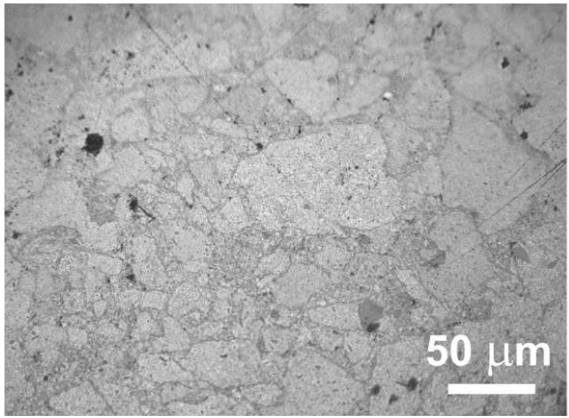
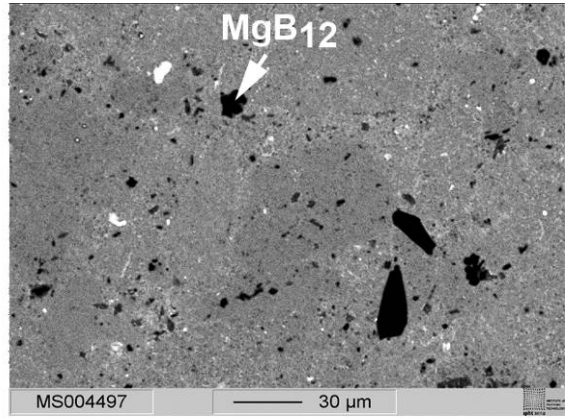
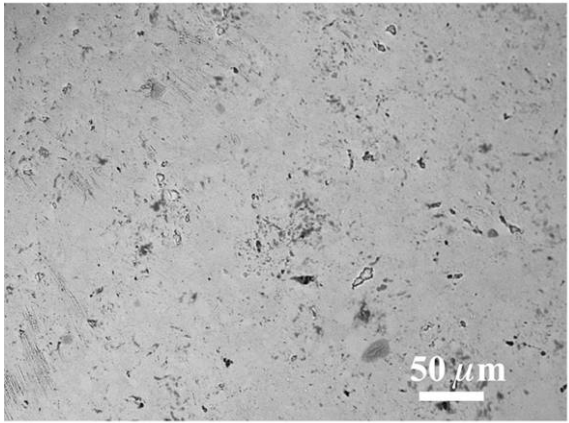
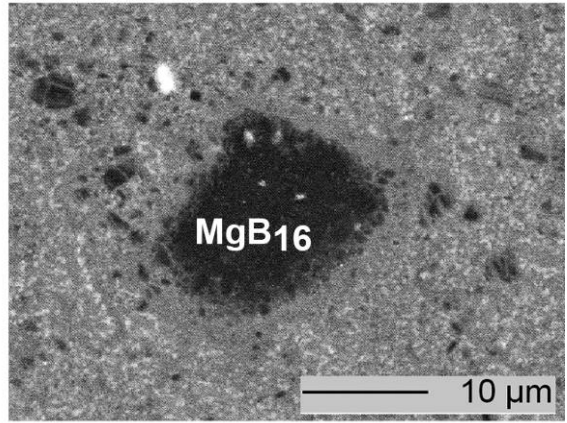

Figure 6.

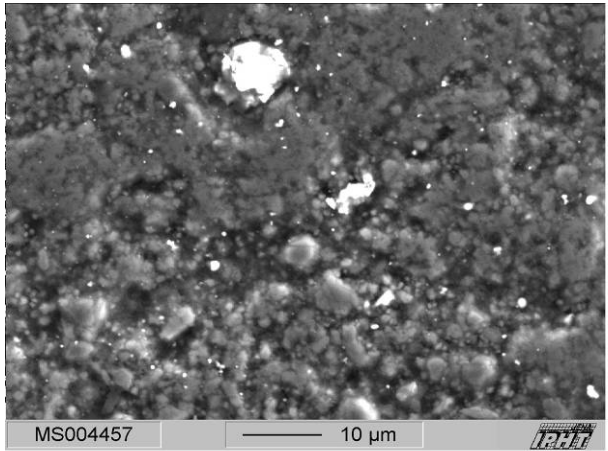 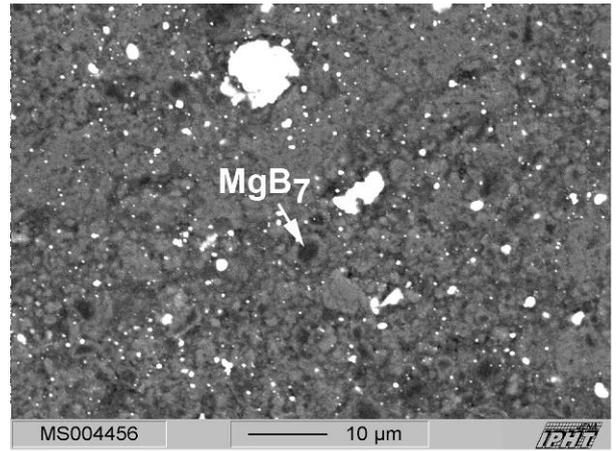
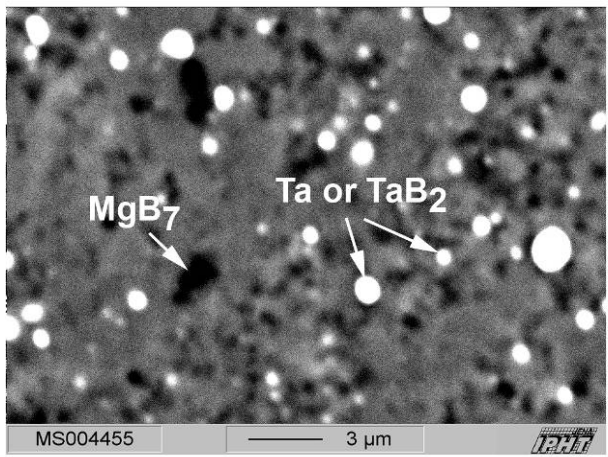 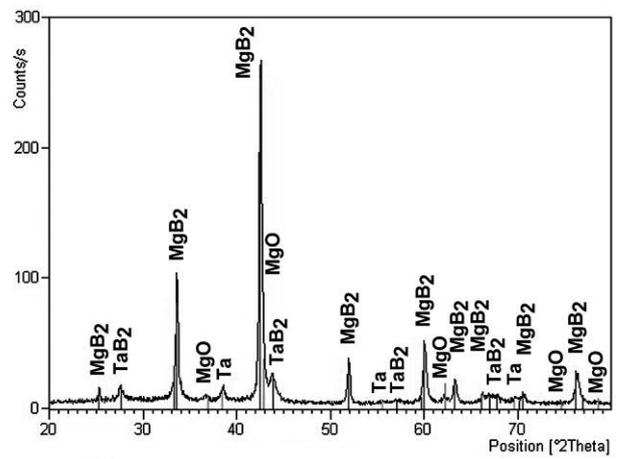

Figure 7.

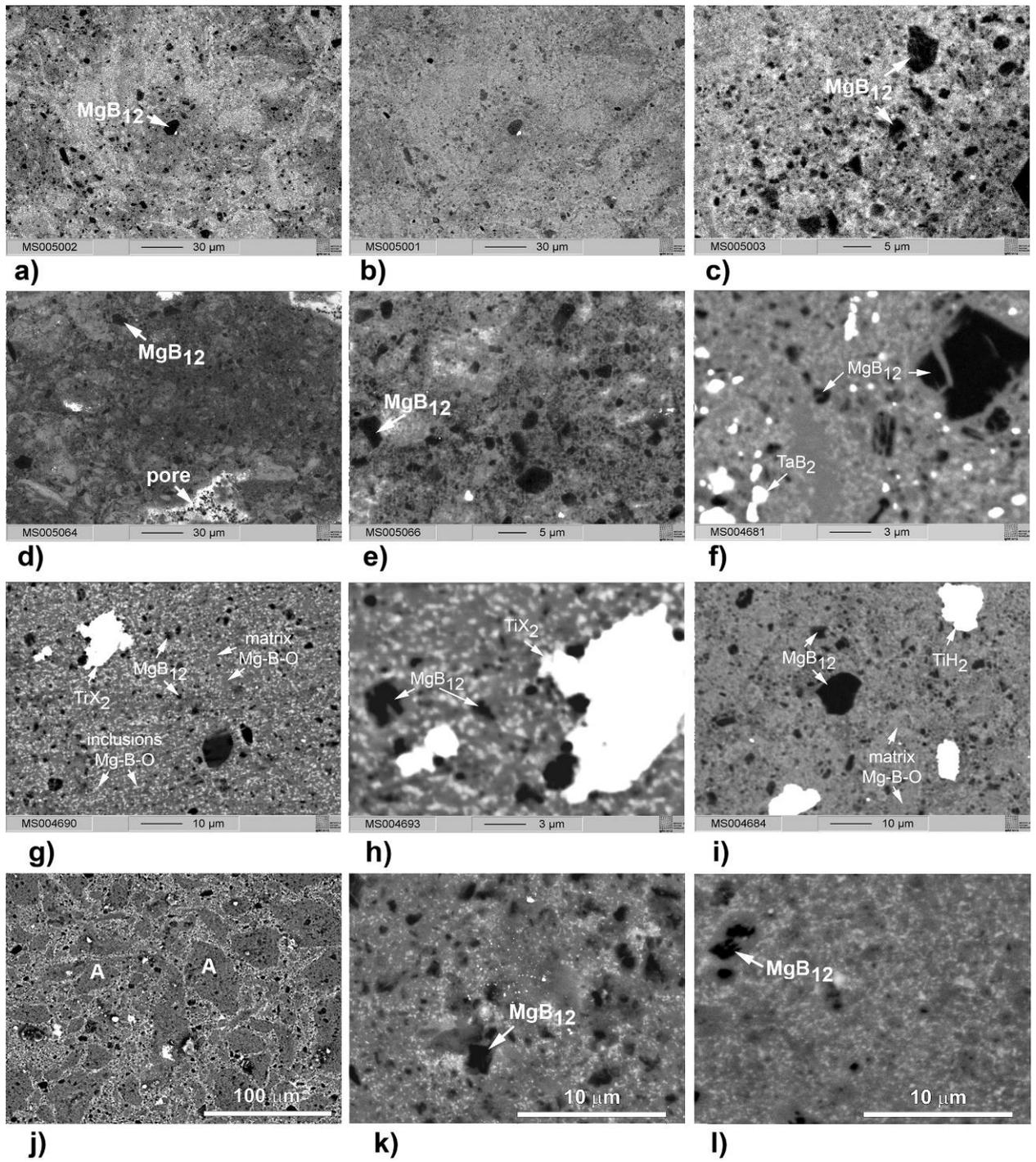

Figure 8.

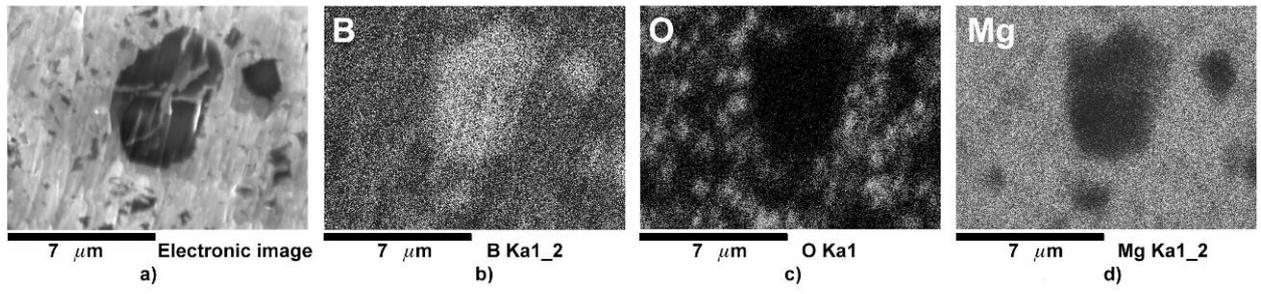

Figure 9.

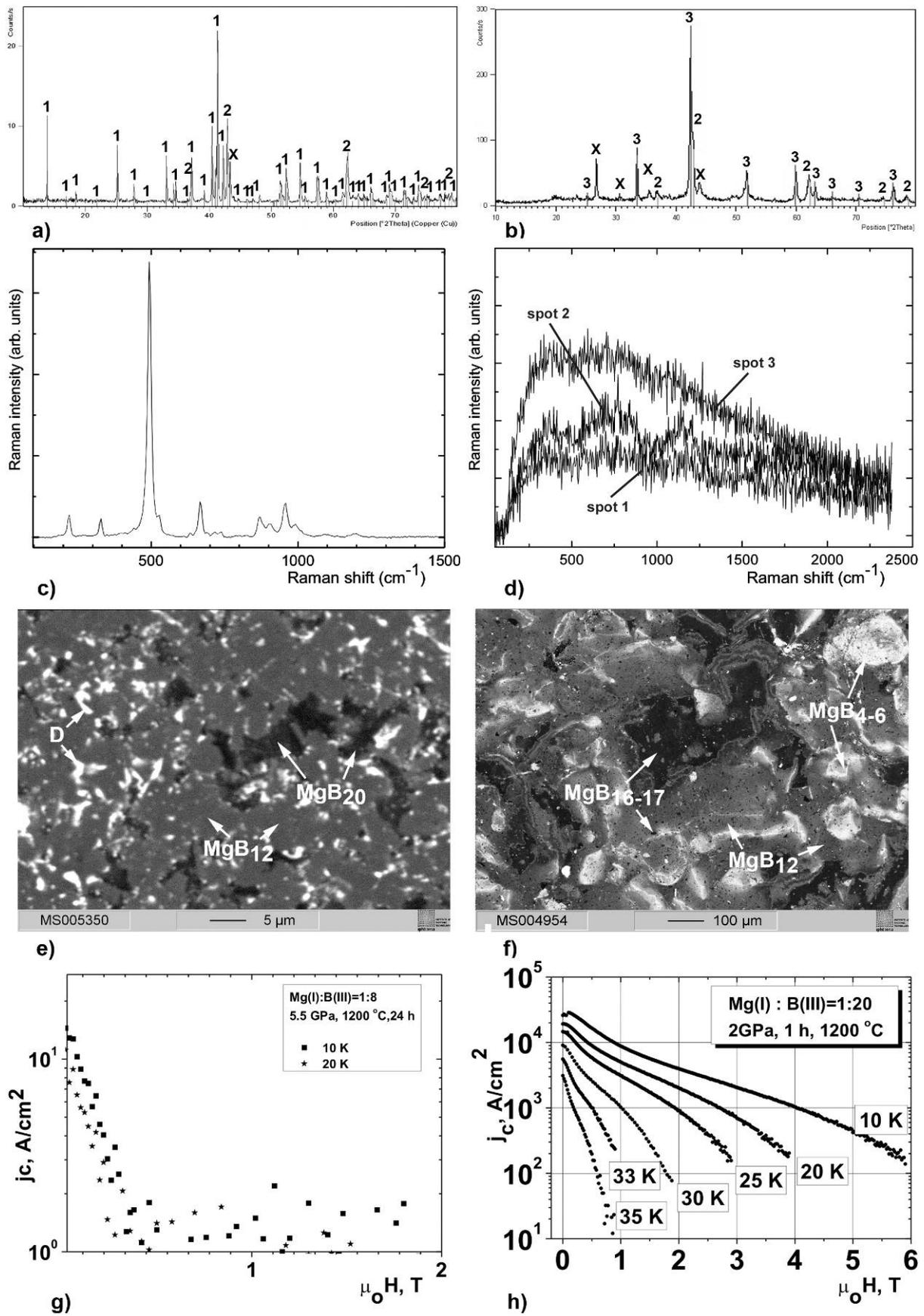

Figure 10

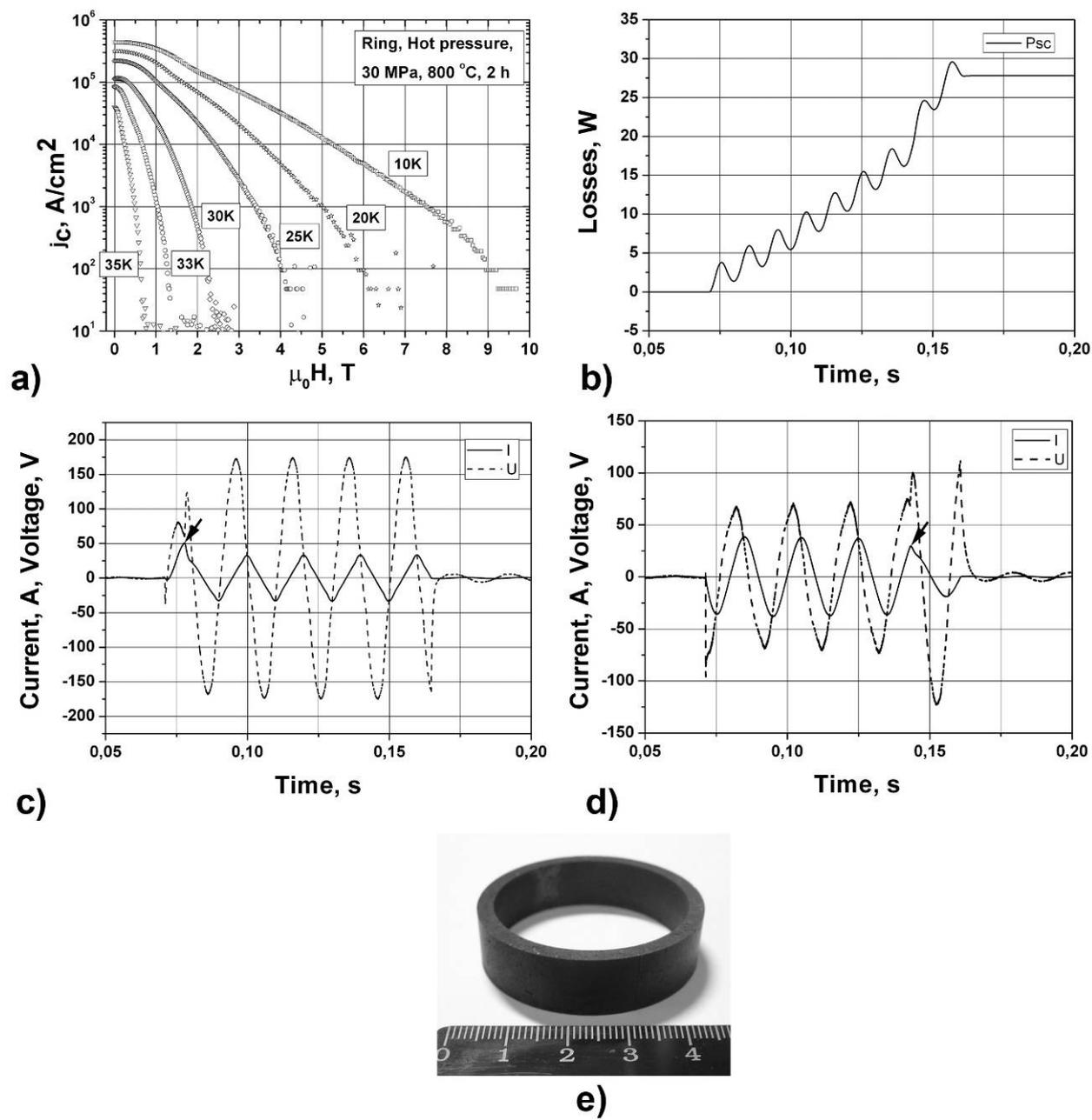

Figure 11